\documentclass[10pt]{article}

\usepackage{parskip}
\usepackage{titlesec}
\usepackage[section]{placeins}
\usepackage{xcolor}
\usepackage{graphicx}
\usepackage{amsfonts,amsmath,amssymb}
\usepackage[utf8]{inputenc}
\usepackage[english]{babel}
\usepackage{float}
\usepackage[margin=1.5in]{geometry}
\usepackage{helvet}
\usepackage{braket}
\usepackage{longtable}
\usepackage{makecell}
\usepackage[super,comma,sort&compress]{natbib}
\usepackage{epstopdf}

\setcitestyle{open={},close={}}

\newcolumntype{X}{p{0.95\textwidth}}

\PassOptionsToPackage{hyphens}{url}

\usepackage[colorlinks = true,
            linkcolor = blue,
            urlcolor  = blue,
            citecolor = blue,
            anchorcolor = blue]{hyperref}

\titleformat{\section}{\bfseries\large}{\thesection.}{0.333em}{}
\titleformat{\subsection}{\bfseries}{\thesubsection}{0.333em}{}
\titleformat{\subsubsection}{\itshape}{\thesubsubsection}{0.333em}{}
\titlespacing{\section}{0pt}{*3}{*1}
\titlespacing{\subsection}{0pt}{*2}{*0.5}
\titlespacing{\subsubsection}{0pt}{*1.5}{0pt}

\begin{document}

\section*{Article Title}

Deep learning quantum Monte Carlo for solids

\section*{Article Category}

Advanced Review

\section*{Authors}

\begin{longtable}{|X|}
\hline
\makecell[X]{
    \textbf{Yubing Qian}*\\
    ORCID iD: 0000-0001-6980-163X\\
    Affiliation:
    \begin{enumerate}
        \item School of Physics, Peking University, Beijing 100871, People’s Republic of China
        \item ByteDance Research, Fangheng Fashion Center, No. 27, North 3rd Ring West Road, Haidian District, Beijing 100098, People’s Republic of China
    \end{enumerate}
    Email: \texttt{phyqyb@pku.edu.cn}\\
} \\
\hline
\makecell[X]{
    \textbf{Xiang Li}*\\
    ORCID iD: 0000-0001-8572-1875\\
    Affiliation:
    \begin{enumerate}
    \item ByteDance Research, Fangheng Fashion Center, No. 27, North 3rd Ring West Road, Haidian District, Beijing 100098, People’s Republic of China
    \end{enumerate}
    Email: \texttt{lixiang.62770689@bytedance.com}
} \\
\hline
\makecell[X]{
    \textbf{Zhe Li}*\\
    ORCID iD: 0000-0002-2493-9229\\
    Affiliation:
    \begin{enumerate}
    \item ByteDance Research, Fangheng Fashion Center, No. 27, North 3rd Ring West Road, Haidian District, Beijing 100098, People’s Republic of China
    \end{enumerate}
    Email: \texttt{lizhe.qc@bytedance.com}
} \\
\hline
\makecell[X]{
    \textbf{Weiluo Ren}*\\
    ORCID iD: 0000-0002-4276-4856\\
    Affiliation:
    \begin{enumerate}
    \item ByteDance Research, Fangheng Fashion Center, No. 27, North 3rd Ring West Road, Haidian District, Beijing 100098, People’s Republic of China
    \end{enumerate}
    Email: \texttt{renweiluo@bytedance.com}
} \\
\hline
\makecell[X]{
    \textbf{Ji Chen}*\\
    ORCID iD: 0000-0003-1603-1963\\
    Affiliation:
    \begin{enumerate}
    \item School of Physics, Peking University, Beijing 100871, People’s Republic of China
    \item Interdisciplinary Institute of Light-Element Quantum Materials, Frontiers Science Center for Nano-Optoelectronics, Peking University, Beijing 100871, People’s Republic of China
    \end{enumerate}
    Email: \texttt{ji.chen@pku.edu.cn}
} \\
\hline
\end{longtable}

\section*{Conflicts of interest}

The authors declare that they have no conflicts of interest.

\section*{Abstract}

Deep learning has deeply changed the paradigms of many research fields.
At the heart of chemical and physical sciences is the accurate \textit{ab initio} calculation of many-body wavefunction, which has become one of the most notable examples to demonstrate the power of deep learning in science.
In particular, the introduction of deep learning into quantum Monte Carlo (QMC) has significantly advanced the frontier of \textit{ab initio} calculation, offering a universal tool to solve the electronic structure of materials and molecules.
Deep learning QMC architectures were initial designed and tested on small molecules, focusing on comparisons with other state-of-the-art \textit{ab initio} methods. 
Methodological developments, including extensions to real solids and periodic models, have been rapidly progressing and reported applications are fast expanding.  
This review covers the theoretical foundation of deep learning QMC for solids, the neural network wavefunction ansatz, and various of other methodological developments.
Applications on computing energy, electron density, electric polarization, force and stress of real solids are also reviewed.
The methods have also been extended to other periodic systems and finite temperature calculations.
The review highlights the potentials and existing challenges of deep learning QMC in materials chemistry and condensed matter physics.

\section*{Graphical/Visual Abstract and Caption}

\begin{center}
    \includegraphics[]{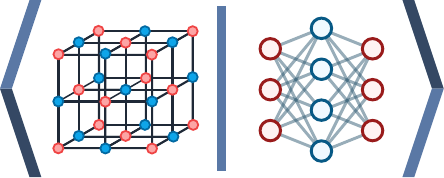}
\end{center}

\textbf{Caption}:
The introduction of deep learning into quantum Monte Carlo has significantly advanced the frontier of \textit{ab initio} calculation.
The method has been extended from molecular systems to real solids and periodic models, showing a great potential for accurate calculations of chemical and physical properties, and exploration of intriguing phenomena.

\newpage

\section{Introduction}

Solving the Schrödinger equation for many-electron systems is an essential task of first-principle modeling, i.e. \textit{ab initio} calculation, of materials and condensed matter systems.
Although \textit{ab initio} calculations have been widely employed now in chemistry and physics studies, 
an exact \textit{ab initio} treatment is impractical and accurate solutions are only available for small enough molecules.
The challenge mainly stems from the many-body electron correlation effects embedded in the Coulomb interactions between many electrons of such systems.
To ease the problem and go forward to obtain a reliable solution of real materials, there are two different directions. 
One is to look for deterministic solutions based on necessary approximations to trade off accuracy for efficiency.
Hartree--Fock (HF) approximation and post Hartree--Fock wavefunction theories such as coupled cluster and configuration interaction can systematically improve the accuracy by explicit inclusion of higher level electron excitation.
The limitation is that the computational cost scales up so rapidly with the number of electrons, which hinders their applications to materials and condensed matter systems.
In a different direction, the density functional theory achieves a good balance between efficiency and accuracy, but the design and choice of the exchange-correlation functional cause questions about the reliability of DFT to strongly correlated systems.
More accurate methods going beyond DFT via e.g. many-body perturbation are also being actively developed, but most of them fall into similar situations that the computational cost scales up rapidly. 
Nowadays, there are many comprehensive reviews and textbooks of these deterministic approaches,~\cite{jensen_introduction_2017,sholl_density_2009,leng_gw_2016} and in this review we focus on the latest development of another type of approach, namely quantum Monte Carlo (QMC), in which the main idea is to look for stochastic solutions to maintain accuracy with the cost of introducing statistical errors. 

An attractive feature of QMC is the possibility of obtaining direct and accurate treatment of many-body correlation effects with a favorable scaling.
The accuracy of QMC can also be systematically improved with an increased number of Monte Carlo samples.
Significant progress has made in the past few decades to expand the capability of QMC, 
e.g. the developments in optimization methods,~\cite{sorella_green-sr_1998,sorella_generalized_2001,neuscamman_optimizing_2012}
interatomic force evaluation,~\cite{filippi_correlated_2000,assaraf_zero-variance_2003,sorella_algorithmic_2010,nakano_space-warp_2022}
and phonon calculation,~\cite{nakano_atomic_2021,ly_phonons_2022} to name a few.
However, there are still some challenges for QMC simulation for materials and condensed matter systems. 
For example, the required computational resource is large, and thus the simulation size is limited, which gives rise to the finite size problem.
Just as importantly, the accuracy in large systems is still to be improved.

Since the seminal review of Foulkes \textit{et al}.\cite{foulkes_solid-review_2001} in 2001, VMC and DMC remains state-of-the-art for accurate calculation of solid materials, with methodological contributions from a diverse community.\cite{kim_qmcpack_2018,needs_variational_2020,nakano_turborvb_2020,wheeler_pyqmc_2023}
Among all the developments, the past few years have witnessed an explosion in the application of deep learning techniques in the development of wavefunction ansatz for VMC.
The neural network takes in Monte Carlo samples of electron coordinates, and outputs the value of the wavefunction, requiring no external data.
Initially, these are applied in spin model systems,~\cite{nomura_restricted_2017,carleo_constructing_2018}
and molecule systems.~\cite{han_deepwf_2019,pfau_ferminet_2020,hermann_paulinet_2020,entwistle_electronic_2023,choo_fermionic_2020,scherbela_deeperwin_2022}
Later, they are generated to real solids~\cite{li_deepsolid_2022,yoshioka_rbm-solid_2021} and other condensed matter systems such as the electron gas~\cite{wilson_wapnet_2023,cassella_discovering_2023,pescia_boson_2022} by encoding periodicity into the neural network.
The employment of neural networks not only improves the accuracy, but also offers new opportunities to re-visit other aspects and techniques used in traditional QMC for materials and condensed matter systems.
In this advanced review, we would like to discuss the recent developments and applications of deep learning QMC for solids.
We note that although the currently working architectures for solids are extended from existing architectures for molecules or lattice models, unique aspects in modeling solids, such as periodicity and finite-size errors, should be clarified.
In addition to the technical differences, the properties of interest go beyond simply energies, and properties like susceptibility, stress tensors, and phonon spectra are also of significant interest.
Therefore, we aim to provide a complementary review article to the seminal review articles of traditional QMC and the latest review on deep learning QMC, which mostly focuses on molecules and lattice models.
The readers are recommended to read the previous reviews ~\cite{foulkes_solid-review_2001,needs_variational_2020} for a comprehensive understanding of traditional QMC, and to read the article of Hermann \textit{et al}.~\cite{hermann_nnqmc-review_2023} for details about different flavors of machine learning architectures for many-body wavefunction ansatz.
In section \ref{sec:methods} we briefly introduce the key aspects of QMC theory and detail the developments of deep learning QMC for solids.
In section \ref{sec:applications} we discuss the applications of deep learning QMC for solids including real solid materials and other condensed matter systems such as electron gas.

\section{Methodological developments}\label{sec:methods}

\subsection{Many-electron Schrödinger equation in solids}

The starting point of QMC and other \textit{ab initio} methods is the time-independent Schr\"odinger equation.
Considering nucleus masses are significantly larger than electrons, it is advantageous to decouple their movements with Born--Oppenheimer approximation, treating the coordinates of the nuclei $\mathbf{R}_I$ as fixed parameters when solving the Schrödinger equation of the electrons:
\begin{gather}
    \hat H \Psi(\mathbf{x}_1,\dots,\mathbf{x}_N) = E \Psi(\mathbf{x}_1,\dots,\mathbf{x}_N),\label{eq:schrodinger}\\
    \hat H = -\frac{1}{2} \sum_{i=1}^N \nabla^2_i
        - \sum_{i=1}^N\sum_{I=1}^{N_{\text{atom}}} \frac{Z_I}{|\mathbf{r}_i-\mathbf{R}_I|}
        + \frac{1}{2} \sum_{i=1}^N\sum_{\substack{j=1\\ j\ne i}}^{N} \frac{1}{|\mathbf{r}_i-\mathbf{r}_j|}
        + \frac{1}{2} \sum_{I=1}^{N_{\text{atom}}}\sum_{\substack{J=1\\ J\ne I}}^{N_{\text{atom}}} \frac{Z_I Z_J}{|\mathbf{R}_I-\mathbf{R}_J|},
\end{gather}
where $Z_I$ is the charge of the nucleus $I$, $\mathbf{x}_i=(\mathbf{r}_i,s_i)$ is the spatial and spin coordinates of the electron $i$, and $\Psi$ is the antisymmetric wavefunction under the permutation of $(\mathbf{x}_1,\dots,\mathbf{x}_N)$.

Given the substantial number of electrons in real solids, exact calculations become intractable, and the supercell approximation is introduced to simplify the problem which enforces the translational symmetry.
A supercell is constructed by tiling multiple primitive cells, and the Coulomb interaction between particles in different supercells is approximated by the interaction between particles and their periodic images in other supercells.
In essence, supercell Hamiltonian can be expressed in the following form
\begin{multline}\label{eq:supercell-hamiltonian}
    \hat H_s = -\frac{1}{2} \sum_{i=1}^N \nabla^2_i
        - \sum_{i=1}^N\sum_{I=1}^{N_{\text{atom}}}\sum_{\mathbf{L}_s} \frac{Z_I}{|\mathbf{r}_i-\mathbf{R}_I+\mathbf{L}_s|} \\
        + \frac{1}{2} \sum_{i=1}^N\sum_{j=1}^{N}\sum_{\mathbf{L}_s}' \frac{1}{|\mathbf{r}_i-\mathbf{r}_j+\mathbf{L}_s|}
        + \frac{1}{2} \sum_{I=1}^{N_{\text{atom}}}\sum_{J=1}^{N_{\text{atom}}} \sum_{\mathbf{L}_s}' \frac{Z_I Z_J}{|\mathbf{R}_I-\mathbf{R}_J+\mathbf{L}_s|}, 
\end{multline}
where $\{\mathbf{L}_p\}$, $\{\mathbf{L}_s\}$ denotes the lattice vectors of primitive cell and supercell respectively. And the summation with a prime means that the $\mathbf{L}_s=0$ term is excluded when $i=j$ or $I=J$.
It should be pointed out that the summation of the Coulomb potentials is conditional convergent, and should be treated specially, usually through the Ewald summation.~\cite{foulkes_solid-review_2001,ewald_berechnung_1921,toukmaji_ewald_1996}

\subsection{Variational Monte Carlo method}

Variational Monte Carlo is a QMC method that employs the variational principle to solve the Schrödinger equation Eq.~\eqref{eq:schrodinger}.
The variational principle asserts that the expectation value of energy $E_v$ for any $\Psi_T$ is always greater or equal to the ground state energy $E_0$:
\begin{equation}
    E_v = \dfrac
        {\int \Psi_T^*(\mathbf{X}) \hat H \Psi_T(\mathbf{X})\:\mathrm{d}\mathbf{X}}
        {\int \Psi_T^*(\mathbf{X}) \Psi_T(\mathbf{X})\:\mathrm{d}\mathbf{X}}
    \ge E_0,
\end{equation}
and the equal sign is taken if and only if $\Psi_T$ is exactly the same as the ground state wavefunction $\Psi_0$.

Given a trial wavefunction $\Psi_T$, the above integration can be calculated using Monte Carlo algorithms.
Specifically, a large number ($M$) of samples $\{\mathbf{X}_i\}$ obeying the probability distribution
\begin{equation}\label{eq:p(X)}
    p_T(\mathbf{X}) = \dfrac
        {\Psi_T^*(\mathbf{X}) \Psi_T(\mathbf{X})}
        {\int \Psi_T^*(\mathbf{X}) \Psi_T(\mathbf{X})\:\mathrm{d}\mathbf{X}}
\end{equation}
can be obtained with the Markov-chain Monte Carlo and the Metropolis--Hastings algorithm.~\cite{foulkes_solid-review_2001,thijssen_computational_1999}
The energy $E_v$ is represented by the average local energy $E_L = [\hat H \Psi_T(\mathbf{X})]/\Psi_T(\mathbf{X})$,
\begin{equation}
    E_v = \left\langle E_L(\mathbf{X})\right\rangle
        = \int p_T(\mathbf{X}) E_L(\mathbf{X}) \:\mathrm{d}\mathbf{X}
        = \lim_{M\to\infty} \frac{1}{M} \sum_{i=0}^{M} E_L(\mathbf{X}_i),
\end{equation}
where $\langle\cdot\rangle$ means the expectation value under the probability distribution $p_T(\mathbf{X})$.

Afterward, the trial wavefunction $\Psi_T$ is optimized variationally.
The Monte Carlo integration and optimization process are repeated until the energy $E_v$ converges to a satisfactory level, and the final wavefunction $\Psi_T$ is considered a good approximation of the ground state wavefunction $\Psi_0$, within the limitation of the chosen ansatz.

\subsection{Wavefunction ansatz}

The choice of an ansatz for the trial wavefunction is of great importance in VMC, as it ultimately determines the achievable accuracy.
In this section, we will discuss the symmetry requirements, as well as various forms of ansatz, including traditional ones and more recent neural network--based ones.

\subsubsection{Symmetry requirements}

Firstly, a valid ansatz for a many-electron wavefunction must satisfy the fundamental requirement of antisymmetry under the exchange of any two electrons, reflecting the fermionic nature of electrons:
\begin{equation}
    \Psi(\dots,\mathbf{x}_i,\dots,\mathbf{x}_j\dots) =
        -\Psi(\dots,\mathbf{x}_j,\dots,\mathbf{x}_i\dots).
\end{equation}

Furthermore, the supercell Hamiltonian $\hat H_s$ is invariant with a simultaneous translation of all electrons by a vector in $\{\mathbf{L}_p\}$ as well as a translation of any electron by a vector in $\{\mathbf{L}_s\}$,
\begin{subequations}
\begin{gather}
    \hat H_s (\mathbf{r}_1+\mathbf{L}_s,\dots,\mathbf{r}_N) = \hat H_s (\mathbf{r}_1,\dots,\mathbf{r}_N), \\
    \hat H_s (\mathbf{r}_1+\mathbf{L}_p,\dots,\mathbf{r}_N+\mathbf{L}_p) = \hat H_s (\mathbf{r}_1,\dots,\mathbf{r}_N),
\end{gather}
\end{subequations}
Correspondingly, wavefunctions can be labeled by two quantum numbers $\mathbf{k}_p,\mathbf{k}_s$ and transform as below
\begin{subequations}
\begin{gather}
    \Psi(\mathbf{r}_1+\mathbf{L}_s,\dots,\mathbf{r}_N) =
        \mathrm{e}^{\mathrm{i}\mathbf{k}_s\cdot\mathbf{L}_s} \Psi(\mathbf{r}_1,\dots,\mathbf{r}_N),\label{eq:supercell-transl}\\
    \Psi(\mathbf{r}_1+\mathbf{L}_p,\dots,\mathbf{r}_N+\mathbf{L}_p) =
        \mathrm{e}^{\mathrm{i}\mathbf{k}_p\cdot\mathbf{L}_p} \Psi(\mathbf{r}_1,\dots,\mathbf{r}_N),\label{eq:system-transl}
\end{gather}
\end{subequations}
where $\mathbf{k}_p,\mathbf{k}_s$ are constrained in the Brillouin zone of primitive cell and supercell respectively.

It is also worth noting that $\hat{H}_s$ possesses spatial group symmetry of solids, and each quantum state will transform according to the specific representations. However, it is often challenging to specify the exact representation of the ground state and associated transform law \textit{a priori}. As a solution, ground state wavefunctions can be built without constraint under spatial group transform, and are ensured to approach the ground truth via energy minimization.

\subsubsection{Traditional ansatz}

Slater--Jastrow ansatz is one of the most widely employed trial wavefunctions in VMC:~\cite{jastrow_many-body_1955, foulkes_solid-review_2001, kim_qmcpack_2018, nakano_turborvb_2020, needs_variational_2020, wheeler_pyqmc_2023}
\begin{equation}\label{eq:periodic-ansatz}
    \Psi^{\text{SJ}}_{\mathbf{k}_s,\mathbf{k}_p}(\mathbf{X}) =
    \mathrm{e}^{J(\mathbf{X})}
    \begin{vmatrix}
        \psi_1(\mathbf{x}_1) & \cdots & \psi_N(\mathbf{x}_1) \\
        \vdots & \ddots & \vdots \\
        \psi_1(\mathbf{x}_N) & \cdots & \psi_N(\mathbf{x}_N) \\
    \end{vmatrix},
\end{equation}
where we define $\mathbf{X}=(\mathbf{x}_1,\dots,\mathbf{x}_N)$, and $\psi_j(\mathbf{x}_i)$ are spin orbitals.
The determinant part is antisymmetric under electron exchange, while the Jastrow factor $\mathrm{e}^{J}$ is symmetric, and is designed to capture the correlations and satisfy the cusp condition.~\cite{kato_cusp_1957}
It is convenient to remove the spin variables in Eq.~\eqref{eq:periodic-ansatz} by rewriting it as~\cite{foulkes_solid-review_2001}
\begin{equation}
    \Psi^{\text{SJ}}_{\mathbf{k}_s,\mathbf{k}_p}(\mathbf{R}) =
    \mathrm{e}^{J(\mathbf{R})}
    \begin{vmatrix}
        \phi_1(\mathbf{r}_1) & \cdots & \phi_{N_\uparrow}(\mathbf{r}_1) \\
        \vdots & \ddots & \vdots \\
        \phi_1(\mathbf{r}_{N_\uparrow}) & \cdots & \phi_{N_\uparrow}(\mathbf{r}_{N_\uparrow}) \\
    \end{vmatrix}\cdot
    \begin{vmatrix}
        \phi_{N_\uparrow+1}(\mathbf{r}_{N_\uparrow+1}) & \cdots & \phi_N(\mathbf{r}_{N_\uparrow+1}) \\
        \vdots & \ddots & \vdots \\
        \phi_{N_\uparrow+1}(\mathbf{r}_N) & \cdots & \phi_N(\mathbf{r}_N) \\
    \end{vmatrix},
\end{equation}
where $\mathbf{R}=(\mathbf{r}_1,\dots,\mathbf{r}_N)$, and the first $N_{\uparrow}$ electrons are assumed to spin up.
And $\phi_j(\mathbf{r}_i)$ are spatial parts of the spin orbitals.

The translational symmetry can be achieved by taking the spatial orbitals $\phi_j(\mathbf{r}_i)$ with a Bloch form:
\begin{equation}\label{eq:phi}
    \phi_j(\mathbf{r}_i) = \mathrm{e}^{\mathrm{i}\mathbf{k}_j\cdot\mathbf{r}_i} u_{j}(\mathbf{r}_i),
\end{equation}
where $u_{\mathbf{k}}$ is invariant with a simultaneous translation of all electrons by a vector in $\{\mathbf{L}_p\}$ as well as a translation of any electron by a vector in $\{\mathbf{L}_s\}$, and $\mathbf{k}_j$ lies on the grid of supercell reciprocal lattice vectors $\{\mathbf{G}_s\}$ offset by ${\mathbf{k}_s}$ within the first Brillouin zone of the primitive cell.

If the determinant part is fixed, the nodes of the wavefunction, namely points where $\Psi(\mathbf{X})=0$, are unchanged during optimization, putting a limitation on the accuracy.
A popular approach is using the backflow transformation,~\cite{kwon_effects_1993, kwon_effects_1998, foulkes_solid-review_2001, lopez_rios_inhomogeneous_2006, kim_qmcpack_2018} replacing the electron coordinates in the determinant with quasiparticle coordinates.
Another approach is to replace the Slater determinant with more general ones, for example Pfaffian~\cite{bajdich_pfaffian_2006,bajdich_pfaffian_2008} or antisymmetrized geminal power wavefunctions,~\cite{casula_geminal_2003,raghav_toward_2023} empowering the expressiveness of the wavefunction ansatz.

\subsubsection{Neural network--based ansatz}

Inspired by the backflow approach, the wavefunction can be constructed from the ground up with expressive neural networks, having a better treatment of electron correlation and obviating the need for Hartree--Fock or DFT starting points.
The key idea is to rewrite Eq.~\eqref{eq:phi} with
\begin{equation}\label{eq:generalized-phi}
    \phi_j(\mathbf{r}_i, \{\mathbf{r}_{/i}\}) = \mathrm{e}^{\mathrm{i}\mathbf{k}_j\cdot\mathbf{r}_i} u_{j}(\mathbf{r}_i, \{\mathbf{r}_{/i}\}),
\end{equation}
where contributions from all other electrons $\{\mathbf{r}_{/i}\}$ are included in a permutation-equivariant way, keeping the wavefunction antisymmetric.
The simplest permutation-equivariant operation for a neural network layer is to pass all individual inputs through the same function.
To incorporate electron-electron correlation, it is also helpful to append the average outputs of the previous layer to the inputs of the next.
That is the essence of FermiNet.~\cite{pfau_ferminet_2020}
Regarding Jastrow factors, however, they are not an essential part of the wavefunction because the neural network-based orbitals are already powerful enough and can well approximate the cusp conditions.

\begin{figure}
\centering
\includegraphics[width=\textwidth]{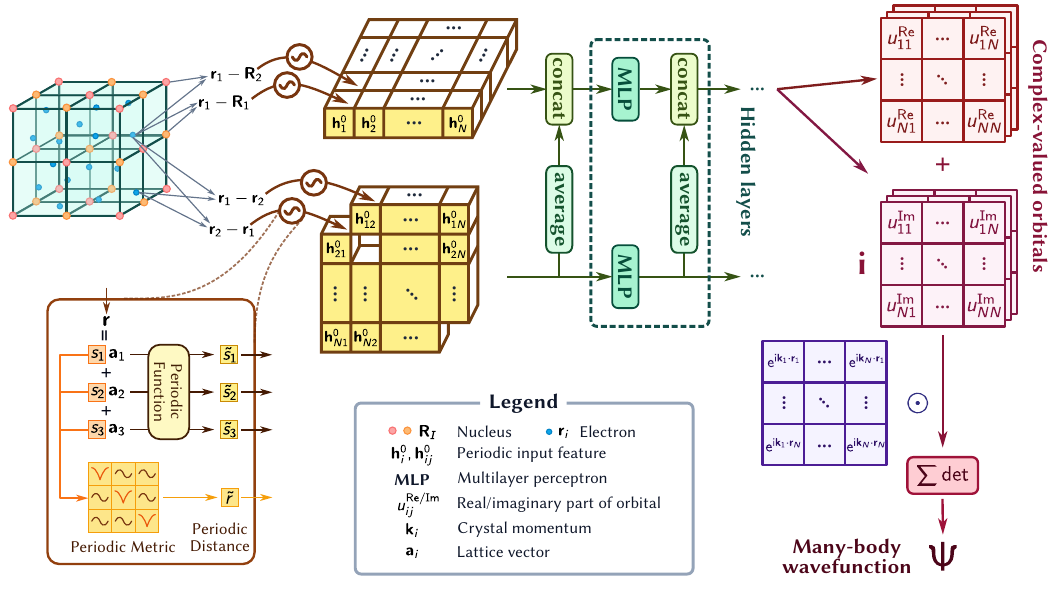}
\caption{
Illustration of the neural network architecture for solids according to DeepSolid.~\cite{li_deepsolid_2022}
Periodicity is imposed on the electron-nucleus and electron-electron distance, forming the input features.
These features are then passed through multiple equivariant hidden layers to construct a set of complex-valued orbitals.
A phase factor is also included.
In the end, the value of the wavefunction is calculated as the weighted summation of the determinants, where each element is a product of a phase factor and a complex-valued orbital.
}
\label{fig:network-structure}
\end{figure}

Fig.~\ref{fig:network-structure} illustrates the architecture of DeepSolid, which is a good example of incorporating periodicity into neural network ansatz.
Neural networks can represent the periodic function $u_{\mathbf{k}}(\mathbf{r})$ by generalizing input coordinate $\mathbf{r}$ and distance $r=|\mathbf{r}|$ to periodic ones $\tilde{\mathbf{r}}$ and $\tilde{r}$.
And it is desirable for $\tilde{r}$ to be proportional to $r$ when $\mathbf{r} \to 0$, reflecting the isotropy of the interactions and ensuring the representation of cusp conditions.
With supercell lattice vectors $\{\mathbf{a}_1,\mathbf{a}_2,\mathbf{a}_3\}$ and corresponding reciprocal lattice vectors $\{\mathbf{b}_1,\mathbf{b}_2,\mathbf{b}_3\}$, an arbitrary vector $\mathbf{r}$ can be decomposed as $\mathbf{r}=s_1\mathbf{a}_1+s_2\mathbf{a}_2+s_3\mathbf{a}_3$, where $s_\alpha = \mathbf{r} \cdot \mathbf{b}_\alpha / (2\pi)$.
The square of the Euclidean distance is then
\begin{equation}
    r^2 =
    \left(\sum_{\alpha=1}^3 \mathbf{a}_\alpha s_\alpha \right)\cdot
    \left(\sum_{\beta=1}^3 \mathbf{a}_\beta s_\beta \right)
    = \sum_{\alpha=1}^3 \sum_{\beta=1}^3 s_\alpha s_\beta \mathbf{a}_\alpha \cdot \mathbf{a}_\beta.
\end{equation}
A typical periodic generalization is to replace $s_\alpha s_\beta$ with $M_{\alpha\beta}(s_\alpha, s_\beta)$:
\begin{equation}
    \tilde{r}^2 = \sum_{\alpha=1}^3 \sum_{\beta=1}^3 M_{\alpha\beta}(s_\alpha, s_\beta) \mathbf{a}_\alpha \cdot \mathbf{a}_\beta,
\end{equation}
where $M_{\alpha\beta}$ is a smooth periodic function satisfying
\begin{equation}
    \lim_{\substack{s_\alpha\to 0\\ s_\beta\to 0}}
    \frac{M_{\alpha\beta}(s_\alpha, s_\beta)}{s_\alpha s_\beta}=\text{constant}.
\end{equation}
Various forms of $M_{\alpha\beta}$ can be chosen, and here we focus on two of them, which we dub ``nu''~\cite{li_deepsolid_2022} and ``sin''~\cite{cassella_discovering_2023}
\begin{subequations}
    \begin{align}
    M_{\alpha\beta}^{\text{nu}}(s_\alpha, s_\beta) &=
        f(s_\alpha)^2\delta_{\alpha\beta} + g(s_\alpha)g(s_\beta)(1-\delta_{\alpha\beta}),\label{eq:nu}\\
    M_{\alpha\beta}^{\text{tri}}(s_\alpha, s_\beta) &=
        [1-\cos(2\pi s_\alpha)][1-\cos(2\pi s_\beta)] + \sin(2\pi s_\alpha)\sin(2\pi s_\beta)\label{eq:tri},
    \end{align}
\end{subequations}
where $f(s)=|s|(1-2|s|^3)$ and $g(s)=s(1-3|s|+2|s|^2)$.
As for the generalized vector $\tilde{\mathbf{r}}$, there is no such restriction, and it can be chosen as $\mathbf{a}_\alpha g(s_\alpha)$,~\cite{li_deepsolid_2022} or a set of $\cos(2\pi s_\alpha)$ and $\sin(2\pi s_\alpha)$.~\cite{cassella_discovering_2023}

Apart from the neural network structure used in FermiNet and DeepSolid, the permutation-equivariant neural network can have many other forms, evolving through an interplay between physics-inspired designs and black-box-like ones. \cite{han_deepwf_2019, pfau_ferminet_2020, hermann_paulinet_2020, li_deepsolid_2022, gerard_gold-standard_2022, gao_pesnet_2022, lin2023explicitly, glehn_psiformer_2023, li_computational_2024}
While most of these advancements have been showcased on molecules and clusters, they can be transferred to solid systems.
Early works utilized sophisticated architecture to outperform traditional methods.  \cite{han_deepwf_2019, pfau_ferminet_2020, hermann_paulinet_2020, gerard_gold-standard_2022}
However, recent transformer-based architectures have demonstrated better performance with an even simpler design. \cite{glehn_psiformer_2023}
A latest development in architecture is the development of LapNet, combined with a new computational framework called forward Laplacian, achieving state-of-the-art accuracy and improving the efficiency of VMC by dozens of times. \cite{li_computational_2024}
Yet, despite these advancements, it remains an intriguing research challenge to design neural networks so that the search space is broad while the optimization landscape is friendly to the training process, especially for systems with strong electron-electron correlations.

\subsection{Wavefunction optimization}

In addition to the expressiveness of the ansatz, it's also important to have the ability to be efficiently optimized with respect to loss function $\mathcal{L}(\boldsymbol{\theta})$.

For traditional ansatz, the stochastic reconfiguration (SR) approach~\cite{sorella_green-sr_1998,sorella_generalized_2001} can efficiently optimize the wavefunction.
Consider a subspace of the Hilbert space spanned by $\{\ket{\Psi_T(\boldsymbol{\theta})},\allowbreak\ket{\partial_{\theta_1}\Psi_T(\boldsymbol{\theta})},\allowbreak\ket{\partial_{\theta_2}\Psi_T(\boldsymbol{\theta})},\allowbreak\dots\}$, where $\ket{\Psi_T(\boldsymbol{\theta})}$ is the trial wavefunction at the current step, and $\theta_i$ are parameters.
Inspired by imaginary time evolution, the operator $(1 - \tau \hat H)$ is applied to the state $\ket{\Psi_T(\boldsymbol{\theta})}$, where $\tau$ is a small imaginary time.
The updated state $\ket{\Psi_T(\boldsymbol{\theta}+\delta\boldsymbol{\theta})}$ is expanded to $x_0 \ket{\Psi_T(\boldsymbol{\theta})} + \sum_j x_j \ket{\partial_{\theta_j}\Psi_T(\boldsymbol{\theta})}$, where generally $x_0 \ne 1$ because the wavefunction is unnormalized.
The coefficients $x_0,x_1,\dots$ are determined by projecting the state $(1 - \tau \hat H)\ket{\Psi_T(\boldsymbol{\theta})}$ back to the subspace:
\begin{subequations}\label{eq:sr-imag-time}
\begin{align}
    \Braket{\Psi_T|(1 - \tau \hat H)|\Psi_T} &= \sum_j \Braket{\Psi_T|\partial_{\theta_j}\Psi_T} x_j + \Braket{\Psi_T|\Psi_T} x_0, \\
    \Braket{\partial_{\theta_i}\Psi_T|(1 - \tau \hat H)|\Psi_T} &= \sum_j \Braket{\partial_{\theta_i}\Psi_T|\partial_{\theta_j}\Psi_T} x_j + \Braket{\partial_{\theta_i}\Psi_T|\Psi_T} x_0,
\end{align}
\end{subequations}
The solution of Eq.~\eqref{eq:sr-imag-time} specifies the parameters should be updated following
\begin{equation}\label{eq:sr}
    \delta \boldsymbol\theta \propto - \mathbf{S}^{-1} \mathbf{f},
\end{equation}
where
\begin{gather}
    S_{ij} = 
        \left\langle
            \frac{\partial \log \Psi_T^*(\mathbf{X})}{\partial \theta_i}\frac{\partial \log \Psi_T(\mathbf{X})}{\partial \theta_j}
        \right\rangle-
        \left\langle\frac{\partial \log \Psi_T^*(\mathbf{X})}{\partial \theta_i}\right\rangle
        \left\langle\frac{\partial \log \Psi_T(\mathbf{X})}{\partial \theta_j}\right\rangle, \label{eq:sr-S}\\
    f_i = \left\langle E_L(\mathbf{X}) \frac{\partial \log \Psi^*_T(\mathbf{X})}{\partial \theta_i} \right\rangle -
        E_v \left\langle \frac{\partial \log \Psi^*_T(\mathbf{X})}{\partial \theta_i} \right\rangle.
\end{gather}

Putting the same things in the language of deep learning, the parameters are optimized in the Riemannian manifold instead of a Euclidean space, and the distance is
\begin{equation}
    \| \mathrm{d} \boldsymbol{\theta} \|^2 = \sum_{i,j} G_{ij}(\boldsymbol{\theta}) \mathrm{d} \theta_i \mathrm{d} \theta_j,
\end{equation}
where $G_{ij}$ is the metric tensor of the space.
Following the steepest descent direction on the Riemannian manifold, we obtain the natural gradient descent scheme:~\cite{amari_natural_1998}
\begin{equation}
    \delta \boldsymbol\theta \propto-\mathbf{G}^{-1} \nabla_{\boldsymbol{\theta}}\mathcal{L}(\boldsymbol{\theta}).
\end{equation}
If the wavefunction is real-valued, we can choose $\mathbf{G}$ as the Fisher information metric $\boldsymbol{\mathcal{F}}$:~\cite{ly_fisher_2017,facchi_qm-fisher_2010}
\begin{equation}
    \mathcal{F}_{ij} = \left\langle
        \frac{\partial \log p(\mathbf{X})}{\partial \theta_i} \frac{\partial \log p(\mathbf{X})}{\partial \theta_j} 
    \right\rangle,
\end{equation}
where $p(\mathbf{X})$ is the normalized probability defined in Eq.~\eqref{eq:p(X)}.
Such a choice is equivalent to SR.~\cite{nomura_restricted_2017, pfau_ferminet_2020}
In the case of a complex-valued wavefunctions, Fubini--Study metric can be utilized, and we take $\mathbf{G}=\mathbf{S}$.~\cite{aniello_classical_2009, facchi_qm-fisher_2010}
Yet, calculating $\mathbf{G}^{-1}$ is impractical for neural networks as the number of parameters is huge.
Instead, approximate $\mathbf{G}^{-1}$ is used with the help of Kronecker-factored approximate curvature (K-FAC).~\cite{martens_optimizing_2015}
Moreover, modifications to K-FAC are needed for handling unnormalized distribution and Fubini--Study metric.~\cite{pfau_ferminet_2020,li_deepsolid_2022}

A few alternative optimization methods are available.
One such approach is the conjugate gradient--based method~\cite{neuscamman_optimizing_2012} which can avoid constructing and storing the large $\mathbf{S}$ matrix, and has been successfully applied to neural network--based ansatz.~\cite{gao_pesnet_2022}
Another notable optimization technique is the linear method, which resembles an approximate Newton method.~\cite{nightingale_optimization_2001, toulouse_optimization_2007, nakano_turborvb_2020}
These strategies together form a diverse toolkit for addressing the challenges of variational wavefunction optimization.

\subsection{Diffusion Monte Carlo}

Besides VMC, Diffusion Monte Carlo is another QMC method that leverages imaginary time projection to obtain accurate ground states.
Given a linear combination of ground and excited states, the imaginary time evolution decays the coefficients of the excited state exponentially, leaving only the ground state.
With the absence of the potential energy term, the imaginary time evolution driven by the kinetic term resembles a diffusion process:
\begin{equation}
    \frac{\partial}{\partial \tau} \Phi(\mathbf{X}, \tau) = \frac{1}{2} \sum_{i=1}^{N} \nabla_i^2 \Phi(\mathbf{X}, \tau).
\end{equation}
To get stable result with the presence of the potential term, a mixed distribution should be used instead of plain $\Psi(\mathbf{X}, \tau)$:
\begin{equation}
    f(\mathbf{X}, \tau)=\Phi(\mathbf{X}, \tau)\Psi_T(\mathbf{X}),
\end{equation}
where $\Psi_T(\mathbf{X})$ is the guiding wavefunction which is often obtained from the VMC process, and further modifications to the diffusion simulation is needed accordingly.~\cite{foulkes_solid-review_2001,umrigar_diffusion_1993,luchow_quantum_2011}

With traditional wavefunction ansatz, due to its limited expressiveness, the VMC process fails to give an accurate-enough wavefunction, and a subsequent DMC process is needed to project out the ground state.
But the DMC process can still produce biased results because of the fixed-node approximation based on inaccurate guiding wavefunction from VMC.
With deep learning VMC, high-quality guiding wavefunctions can be obtained for DMC simulation, and a higher accuracy in energy can be reached.~\cite{ren_towards_2023}
Nevertheless, since the wavefunction from deep learning VMC is already accurate-enough, and many physical quantities, for example, the force~\cite{assaraf_zero-variance_2003} and susceptibility,~\cite{umari_dielectric_2005} are harder to evaluate in DMC as opposed to VMC, the DMC process is not always necessary for deep learning VMC.

\subsection{Pseudopotential}

Up to this point, our discussions have focused on all-electron QMC simulations.
However, for systems involving heavy atoms, this can be inefficient.
The computational cost scales rapidly with the number of electrons. The core electrons have a negligible impact on the properties of the solid, but they contribute a large part of the energy, hence the most computational effort is spent on capturing the complicated behavior of the core electrons.
Pseudopotential, also known as effective core potential (ECP), is extremely useful for the simulation of heavy atoms.~\cite{annaberdiyev_ccecp_2018}
It replaces the core electrons with an effective potential which mimics their influence on valence electrons.
The reduction of the number of electrons enables the simulation of larger systems.
It also benefits the convergence of the neural network by removing the overwhelming energy contributions of the core electron.
Therefore, different kinds of pseudopotentials have been developed for QMC during the past decades, including Burkatzki-Filippi-Dolg (BFD) ECP.~\cite{burkatzki_bfd_2007} and correlation consistent (cc) ECP~\cite{annaberdiyev_ccecp_2018}
The pseudopotential technique has been implemented into neural network QMC in ref.~\citenum{li_fermionic_2022} to study transitional metal atoms and oxides.

\subsection{Finite-size error}\label{sec:finite-size}

The supercell approximation used in calculations for solid systems results in the finite-size error (FSE) concerning the thermodynamic limit.
To reduce FSE on the energy, the most straightforward approach is to perform calculations with different system sizes and extrapolate the energy to the infinite-size limit.~\cite{drummond_finite-size_2008}
However, this approach is usually inapplicable due to large computational costs, hence several more efficient methods to correct FSE have been proposed.

One of the most popular techniques is twist averaging,~\cite{lin_twist-averaged_2001} which reduces FSE while retaining high computational efficiency. 
Its main point is to average results of different twists $\mathbf{k}_s$, which ensures a more thorough sampling of the supercell Brillouin zone and therefore faster convergence towards the thermodynamic limit.
Specifically, the twist averaged expectation value of the observable $\hat O$ reads:
\begin{equation}
    \langle \hat O \rangle_{\text{twist}} = \frac{\Omega_s}{(2\pi)^3}\int \mathrm{d}\mathbf{k}_s\braket{\Psi_{\mathbf{k}_s}|\hat O|\Psi_{\mathbf{k}_s}} \approx \frac{1}{N_{\text{twist}}} \sum_{\mathbf{k}_s} \braket{\Psi_{\mathbf{k}_s}|\hat O|\Psi_{\mathbf{k}_s}},
    \label{eq:tabc}
\end{equation}
where $\Omega_s$ is the supercell volume, and the integral is practically approximated using a discrete sum of the Monkhorst-Pack $k$-point grid.~\cite{monkhorst_special_1976}
Compared with simply increasing supercell size, twist average shows a much more efficient linear complexity to reduce FSE. To further reduce the cost of reaching the thermodynamic limit, there are many attempts to approximate the twist averaged integral in Eq.~\eqref{eq:tabc} with the result of a single special point $\mathbf{k}_{0}$, which reads 
\begin{equation}
     \langle \hat O \rangle_{\text{twist}} \approx \braket{\Psi_{\mathbf{k}_{0}}|\hat O|\Psi_{\mathbf{k}_{0}}}.
\end{equation}
Baldereschi proposed some special points according to the spatial symmetry of solids and achieved better TDL convergence in insulators. \cite{baldereschi_mean-value_1973} 
Later, more special points were proposed by different groups \cite{rajagopal_quantum_1994,dagrada_exact_2016, mihm_shortcut_2021} from analyzing mean-field level results, which further broaden the applicable range of the single-point method. 

Apart from the twist averaging method, the structure factor--correction scheme can be further employed to reduce the FSE in Coulomb interactions. \cite{chiesa_finite-size_2006}
The main contribution of FSE in Coulomb energy origins from the difference between the Ewald summation and the original integral, and can be removed via
\begin{gather}
    \Delta V_N = N \frac{2\pi}{\Omega_s}\lim_{k\to0}\frac{S(\mathbf{k})}{\mathbf{k}^2},\nonumber\\
    S(\mathbf{k})=\frac{1}{N}\left[\langle\rho(\mathbf{k})\rho^*(\mathbf{k})\rangle-\langle\rho(\mathbf{k})\rangle\langle\rho^*(\mathbf{k})\rangle\right], \rho(\mathbf{k})=\sum_i \exp(i\mathbf{k}\cdot\mathbf{r}_i),
\end{gather}
where $N$ is the number of particles and $S(\mathbf{k})$ is the structure factor. In practical calculations, $S(\mathbf{k})$ is usually calculated with finite supercell, the $k\to0$ limit is estimated via extrapolation.

\section{Applications}\label{sec:applications}

Having established the fundamental framework of deep learning QMC methods, we now turn our attention to a review of the simulation results obtained using these techniques in various systems.

\subsection{Real solids}

\subsubsection{Energy}

\begin{figure}
\centering
\includegraphics[width=\textwidth]{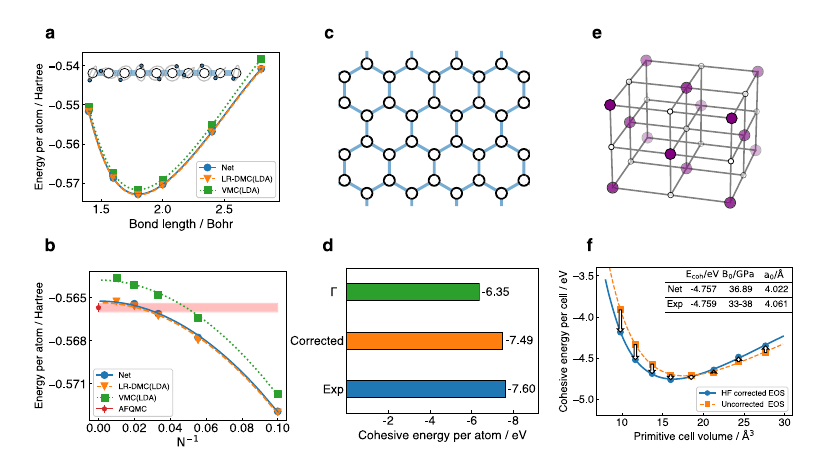}
\caption{%
Energy results for different systems from DeepSolid~\cite{li_deepsolid_2022} (labeled as ``Net'') and other methods.
\textbf{(a)} Hydrogen chain dissociation curve.
\textbf{(b)} Hydrogen chain energy of different size $N$, extrapolated to TDL.
The bond length is fixed at $1.8$ Bohr.
LR-DMC and VMC results~\cite{motta_hydrogen-chain_2017} use the cc-pVTZ basis and DFT orbitals with the local density approximation (LDA) functional.
The AFQMC result~\cite{motta_hydrogen-chain_2017} is extrapolated to the complete basis set limit, and is considered as the current state-of-the-art.
\textbf{(c)} Structure of graphene. The equilibrium length is $1.421$ \r{A}.
\textbf{(d)} Graphene cohesive energy per atom of $\Gamma$ point and finite-size error corrected result.
The experimental data are taken from ref.~\citenum{dappe_local-orbital_2006}.
\textbf{(e)} Structure of rock-salt lithium hydride crystal. The depth of color represents the distance of the points.
\textbf{(f)} Equation of state of lithium hydride crystal, fitted with the Birch--Murnaghan equation of state.~\cite{birch_finite_1947}
The experimental data are taken from ref.~\citenum{nolan_calculation_2009}.
Hartree--Fock based corrections are applied to correct the finite-size error.
Adapted from ref.~\citenum{li_deepsolid_2022},
\href{https://creativecommons.org/licenses/by/4.0/}{CC BY 4.0}.%
}
\label{fig:li_deepsolid_2022}
\end{figure}

The cohesive energy and the equation of state are fundamental properties of solids.
They can be measured in experiments, and results with a wide range of accurate \textit{ab initio} methods are also available, making them perfect benchmarks for new computational methods.
Employing deep learning QMC, Li \textit{et al.}~\cite{li_deepsolid_2022} carried out calculations on various systems, including hydrogen chain, graphene, and lithium hydride crystal, as shown in Fig.~\ref{fig:li_deepsolid_2022}.
For the hydrogen chain, the equation of state from neural networks nearly coincided with the accurate lattice-regularized diffusion Monte Carlo (LR-DMC)~\cite{casula_diffusion_2005} results, and significantly outperformed traditional VMC.
The cohesive energy after extrapolating to the thermodynamic limit (TDL) was also comparable to the state-of-the-art results from LR-DMC and auxiliary field quantum Monte Carlo (AFQMC).~\cite{motta_afqmc_2018}
However, the finite-size error was significant with a limited supercell size when comparing the energy results on graphene and lithium hydride with the experimental data.
The authors corrected the energies with the twist averaged boundary condition (TABC) in conjunction with the structure factor correction,~\cite{chiesa_finite-size_2006} and then the cohesive energy and equation of state were in excellent agreement with the experimental results.

Apart from the finite-size error, the convergence error also plays an important part in relative energies.
Fu \textit{et al.}~\cite{fu_variance_2024} proposed an empirical scheme using the training data to extrapolate energies to zero-variance.
By employing the scheme based on the original DeepSolid training data, the cohesive energy of the hydrogen chain and the graphene were greatly improved.

\subsubsection{Electron density}

Electron density is also a crucial property in solid systems, characterizing features of distinct solids.
Li \textit{et al.}~\cite{li_deepsolid_2022} calculated the electron density of the diamond-structured silicon and rock-salt sodium chloride, which are the typical covalent and ionic crystals, and successfully revealed their characters.
Besides, electron density also indicates whether the solid is a metal or insulator.
As an example, electron densities of hydrogen chains with various bond lengths were calculated by Li \textit{et al.}~\cite{li_deepsolid_2022} with deep learning QMC.
The calculated electron density is rather uniform in a compressed hydrogen chain, and becomes much more localized for a stretched chain, indicating the existence of the metal--insulator transition.

\subsubsection{Electric polarization}

Dielectric response, as one of the fundamental properties of materials, is instrumental in many electromagnetic phenomena.
However, a proper microscopic theory of polarization was not established until the 1990s, which is called the modern theory of polarization.~\cite{king-smith_mtp_1993, resta_macroscopic_1994, resta_theory_2007}
Within this theoretical framework, the electric polarization is closely related to the Berry phase.
However, the currently most used methods such as DFT and Hartree--Fock neglects electron correlations, leading to considerable errors in susceptibility results.~\cite{kirtman_electric_2011}
Although a self-consistent DMC workflow for susceptibility was proposed by Umari \textit{et al.}~\cite{umari_dielectric_2005, umari_linear_2009}, its computational complexity hindered further application.

In contrast, within the deep learning VMC framework, an accurate calculation of dielectric properties becomes feasible. 
Given a finite electric field $\boldsymbol{\mathcal{E}}$, the parameters of the wavefunction are optimized against the electric enthalpy $F$:~\cite{souza_first-principles_2002, umari_dielectric_2005}
\begin{gather}
    F = E_v - \Omega_s \boldsymbol{\mathcal{E}} \cdot \mathbf{P},\nonumber\\
    \mathbf{P}=-\frac{1}{\Omega_s}\sum_{\alpha}\frac{\mathbf{a}_\alpha}{2\pi}
        {\rm Im}\ln\left\langle\exp\left(
        \mathrm{i}\mathbf{b}_\alpha\cdot\sum_i \mathbf{r}_i
        \right)\right\rangle,
\end{gather}
where $\mathbf{P}$ is the polarization density calculated with the modern theory of polarization.

Li \textit{et al.}~\cite{li_electric_2024} applied deep learning QMC to dielectric constant calculations, where neural networks played an important role in incorporating electron correlations and enhance the expressiveness of the wavefunction ansatz.
The accuracy of deep learning QMC was demonstrated on various systems by comparing with the most accurate methods and experimental data. 
For example, the dielectric constant results for alkali metal hydrides are listed in Table.~\ref{tab:XH_dielectric}, and results from DeepSolid are very close to the experimental results.

\begin{table}[H]
\caption{%
High-frequency dielectric constants $\epsilon_\infty$ of alkali metal hydrides.
HF and DeepSolid results are calculated in ref.~\citenum{li_electric_2024}.
DFT results with LDA and Perdew--Burke--Ernzerhof (PBE) functionals come from ref.~\citenum{barrera_lda_2005}, and experimental data come from ref.~\citenum{staritzky_crystallographic_1956, ghandehari_band_1995, batsanov_refractive_2016}.%
}
\label{tab:XH_dielectric}
\begin{center}
\begin{tabular}{l|ccccc}
    \hline
    System & LDA & PBE & HF  & DeepSolid & Experiment \\
    \hline
    $\mathrm{LiH}$ & 4.92 & 4.28 & 3.03 & 3.391(2) & 3.61, 3.939 \\
    $\mathrm{NaH}$ & 3.35 & 3.12 & 2.37 & 2.556(1) & 2.161 \\
    $\mathrm{KH}$ & 2.94 & 2.69 & 2.21 & 1.967(1) & 2.111 \\
    $\mathrm{RbH}$ & 3.07 & 2.73 & 2.25 & 1.818(1) & Not Available \\
    $\mathrm{CsH}$ & 3.45 & 2.98 & 2.41 & 1.695(1) & 1.638 \\
    \hline
\end{tabular}
\end{center}
\end{table}

\subsubsection{Force and stress}

For further application of deep learning QMC to solid systems, the interatomic force and stress tensor are essential physical quantities that are needed for structural optimization, phonon calculation, molecular dynamics, etc.
However, for interatomic force, a straightforward VMC evaluation will lead to an infinite variance.~\cite{hammond_monte_1994,assaraf_zero-variance_2003}
Over the past decades, the infinite variance problem has motivated many researchers developing better force estimators in traditional VMC.~\cite{filippi_correlated_2000, assaraf_zero-variance_2003,chiesa_accurate_2005, badinski_methods_2010}
Qian \textit{et al.}~\cite{qian_interatomic_2022, qian_force_2024} implemented various force estimators and tested their accuracy and efficiency for both solid and molecular systems.
Compared with traditional VMC, force calculations benefited from the quality of the neural network--based wavefunction, demonstrating the power of deep learning QMC by providing a better wavefunction.

Traditionally, the force estimators are often designed with DMC in mind.
Consequently, designing a force estimator specially for deep learning VMC can potentially accelerate calculation and improve accuracy.
Qian \textit{et al.}~\cite{qian_force_2024} proposed a new force estimator called fast-warp estimator, which is a modification based on the space warp coordinate transformation (SWCT) estimator.~\cite{umrigar_two_1989, filippi_correlated_2000, sorella_algorithmic_2010, nakano_space-warp_2022}
Within the deep learning QMC framework, the fast-warp estimator produced smaller variance while being more computationally efficient compared with the SWCT estimator, making it the preferred estimator for force calculation.
The authors also analyzed the performance of periodic input features for force calculations, showing unstable results of the SWCT estimator and highlighting the advantage of the fast-warp estimator.

The calculation of stress tensor in VMC does not suffer from the infinite variance problem, and the underlying quantum mechanics theory is well established.~\cite{nielsen_stress_1985, martin_electronic_2020}
Qian \textit{et al.}~\cite{qian_force_2024} implemented the stress tensor estimator and tested it on the lithium hydride crystal.
The diagonal component of the stress tensor fits very well with the equation of state curve, validating the stress calculation.

\subsection{Other periodic systems}

\subsubsection{Superfluid}

Superconductivity and superfluidity are two famous microscopic quantum phenomena and are intricately interconnected like two sides of one coin.
Both phenomena can be understood with regard to condensation of bosons or paired fermions.
In these exotic phases, couples of fermions are bound together by attractive interactions, which are absent in normal solids.
The weak pairing regime is known as the Bardeen-Cooper-Schrieffer (BCS) state, associated with superconductivity, while the Bose-Einstein condensate (BEC) represents the state of tightly bound fermion pairs, characterizing superfluidity. 
The BCS-BEC crossover, a transition followed by increasing interaction strength, illustrates the continuity within these two limits.
The unitary Fermi gas (UFG) system modulated by an adjustable interaction serves as an ideal platform to simulate BCS-BEC crossover both experimentally and numerically, which has drawn a lot of attention.

Recently, the deep learning QMC was applied to the unitary Fermi gas (UFG) system \cite{pfau_superfluid, kim_superfluid}.
The Hamiltonian for this system incorporates a short-range, strongly attractive interaction between particles of opposite spins:
\begin{equation}
    \hat{H} = -\frac{1}{2} \sum_{i}^{N} \nabla_i^2 + \sum_{ij}^{N / 2}U(\boldsymbol{r}_i^{\uparrow} - \boldsymbol{r}_j^{\downarrow}),
\end{equation}
where $U$ is the modified P\"{o}schl-Teller interaction $U(\boldsymbol{r}) = - 2 \nu_0 \mu^2 / \cosh^2(\mu r)$.
To more accurately capture the electronic structure of the UFG system, wave function ansatzes incorporating pairing features, such as the antisymmetric geminal power singlet (AGPs) \cite{pfau_superfluid} and a more generalized Pfaffian form \cite{kim_superfluid}, were employed:
\begin{equation}
    \Psi_{\rm AGPs(Pfaffian)}=\operatorname{Pf}\left[\phi(\mathbf{r}_i^{\uparrow (\uparrow, \downarrow)}, \mathbf{r}_j^{\downarrow (\uparrow, \downarrow)})\right],
\end{equation}
where AGPs specifically pairs electrons of opposite spins, whereas the Pfaffian accounts for the pairing of any two electrons.
The emergence of pairing was suggested by gap analysis for these pairing-structured ansatzes.
In contrast, the Slater-Jastrow type ansatz, even when augmented with backflow, did not adequately capture such correlations.
Furthermore, the specially designed ansatz combined with deep learning achieved the lowest energy and outperformed the state-of-the-art results from fixed node diffusion Monte Carlo.

\subsubsection{Electron gas}

Homogeneous electron gas (HEG) is one of the simplest model systems with electron--electron correlation in condensed matter physics, yet has proven to exhibit a wide range of phenomena.
Different groups, such as Wilson \textit{et al.}~\cite{wilson_wapnet_2023}, Cassella \textit{et al.}~\cite{cassella_discovering_2023}, Li \textit{et al.}~\cite{li_deepsolid_2022}, and Pescia \textit{et al.},~\cite{pescia_message-passing_2023} applied deep learning QMC to study the energy of HEG.
Wilson \textit{et al.}~\cite{wilson_wapnet_2023} found that by adding a backflow process where the coordinated are generated from the neural network, the energy can be significantly lowered, and is comparable or better than the fixed-node iterative backflow DMC (IB-DMC) results.
Pescia \textit{et al.}~\cite{pescia_message-passing_2023} built a neural network--based backflow ansatz with orders of magnitudes fewer parameters and carried out calculation with $N=128$ electrons, enabling future work on finite-size extrapolations to the thermodynamic limit.

Besides the energy calculation, the HEG system is of particular interest for hosting Wigner crystal phases. As the electron density decreases, HEG systems are expected to transition from a Fermi-liquid phase to a Wigner crystal phase, where electrons spontaneously organize into crystals. Despite this phenomenon being predicted years ago, accurately locating the critical density for this phase transition remains a significant challenge. 
In traditional Quantum Monte Carlo (QMC) methods, different ansatzes are typically constructed for the Fermi-liquid and Wigner crystal phases based on \textit{a priori} knowledge, and the ground state is determined by comparing the energies of these different ansatzes. 
However, with the flexibility of neural networks, a single form of neural network-based ansatz can capture different phases, providing a unified description of the phase transition and yielding more accurate results.

\begin{figure}
\centering
\includegraphics[width=\textwidth]{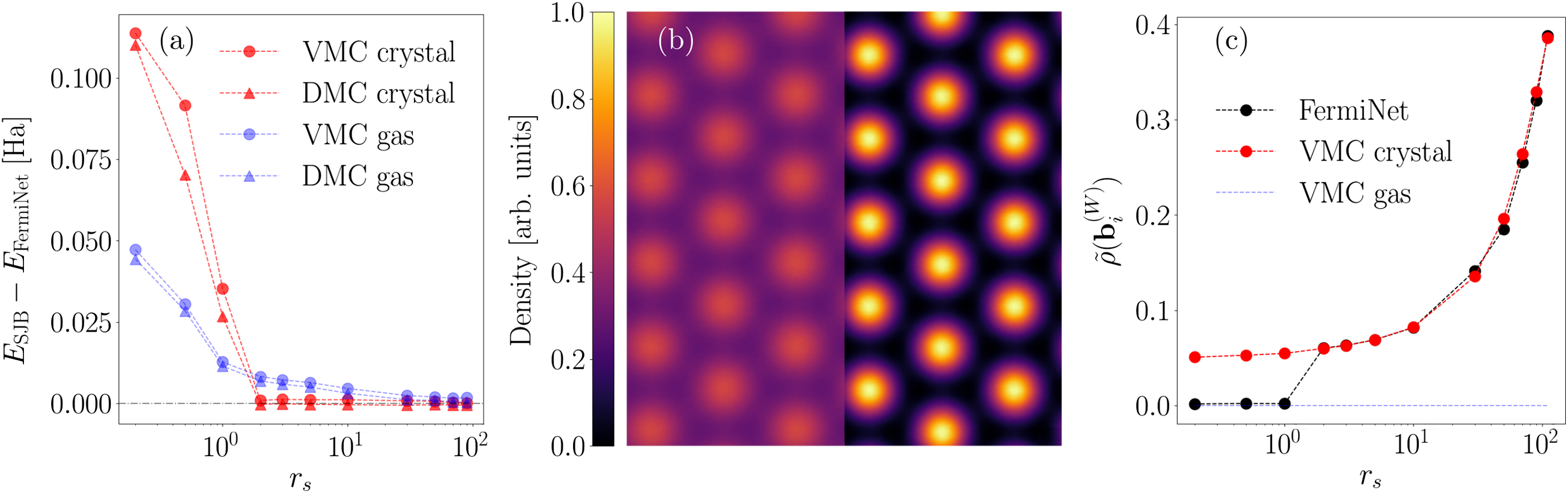}
\caption{%
$N=27$ spin polarized HEG.
\textbf{(a)} Plane wave (gas) and Gaussian orbital (crystal) single-determinant Slater--Jastrow backflow (SJB) ground-state energies per electron relative to FermiNet results.
\textbf{(b)} One-electron density results from FermiNet with $r_s = 10$ (left) and $r_s=70$ (right).
\textbf{(c)} Order parameter for the Wigner crystal state, averaged over crystal axes.
The order parameter rises sharply to a finite value at $r_s=2$, corresponding to the emergence of a crystalline state.
Reprinted from ref.~\citenum{cassella_discovering_2023},
\href{https://creativecommons.org/licenses/by/4.0/}{CC BY 4.0}.%
}
\label{fig:cassella_discovering_2023}
\end{figure}

Specifically, Cassella \textit{et al.}~\cite{cassella_discovering_2023} extended FermiNet with periodic input features Eq.~\eqref{eq:tri}, and allowed representation of symmetry broken Wigner crystal state by rewriting Eq.~\eqref{eq:generalized-phi} with
\begin{equation}
    \phi_j(\mathbf{r}_i, \{\mathbf{r}_{/i}\}) = \sum_m \nu_{jm} \mathrm{e}^{\mathrm{i}\mathbf{k}_m\cdot\mathbf{r}_i} u_{j}(\mathbf{r}_i, \{\mathbf{r}_{/i}\}),
\end{equation}
where $\mathbf{k}_j$ are supercell reciprocal lattice vectors up to the Fermi wave vector of the noninteracting electron gas, and $\nu_{jm}$ are learnable parameters.
Thus, the symmetry Eq.~\eqref{eq:system-transl} can be broken, while symmetry Eq.~\eqref{eq:supercell-transl} is still satisfied to keep the ansatz valid.
The neural network was applied to 27-electron spin-polarized HEG with different densities, and the results are shown in Fig.~\ref{fig:cassella_discovering_2023}.
The energy accuracy of the neural network was very close to traditional DMC using single-determinant Slater--Jastrow with backflow.
By calculating the order parameter, i.e. the Fourier component of the one-electron density, they found that the neural network was capable of expressing both the Fermi liquid phase and the Wigner crystal phase, with the phase transition occurring around $r_s=2$.
Overall, neural network wavefunctions show great potential in discovering novel quantum phases and other emergent phenomena in condensed matter.

\subsection{Dense hydrogen at finite temperature}

Besides solving electronic properties under zero temperature, deep learning QMC can also be used in conjunction with other methods to compute finite temperature properties.
Although molecular dynamics is one of the most commonly used methods to calculate the properties under finite temperatures, we have to first converge the electron wavefunction and get the Born--Oppenheimer potential energy surface before moving the nucleus, which is very expensive.

Xie \textit{et al.}~\cite{xie_deep_2023} developed a neural network--based variational free energy approach for dense hydrogen, using a neural network to represent the wavefunction under a specific proton configuration, and another normalizing flow network to model the proton distribution.
These two neural networks are optimized simultaneously by minimizing free energy,
\begin{equation}
    F=\mathop{\mathbb{E}}_{\mathbf{S}\sim p(\mathbf{S})} \left[
    k_B T \log p(\mathbf{S}) + E_v(\mathbf{S})
    \right],
\end{equation}
where $E_v(\mathbf{S})$ stands for the energy at given proton coordinates $\mathbf{S}$, and the overall expectation is taken under the proton probability density $p(\mathbf{S})$.
With this framework, the free energy, energy, entropy, and pressure of high-temperature dense hydrogen can be calculated with minimal physical constraints.

\section*{Conclusion}

Neural networks can serve as a universal, powerful and flexible ansatz for QMC, offering accurate treatment of electron correlations.
By obtaining high-quality many-body wavefunction within the VMC framework, the neural network enables accurate QMC calculations of many observables that are otherwise difficult to compute in projection-based QMC approaches.
With the flexibility of the neural network, wavefunction ansatz can be constructed without \textit{a priori} knowledge of the ground state, enabling future studies of many interesting phenomena, including a diversity of phase transitions, quantum Hall physics, topological materials, etc.

Going forward, it is necessary to further reduce the high computational cost that is currently limiting the application of neural network QMC.
Employing pseudopotentials is a natural option, but the computational cost arising from the nonlocal integration is still intractable.
One potential solution is to transform the nonlocal potentials into local pseudo-Hamiltonian by modifying the kinetic energy operator.~\cite{bachelet_novel_1989, foulkes_pseudopotentials_1990}
Although this may result in a loss of accuracy, it is still acceptable when compared to the efficiency gain.
In addition, applying more sophisticated and efficient optimization schemes, such as the Wasserstein QMC approach~\cite{neklyudov_wasserstein_2023}, offers potential acceleration in convergence and reduction in computational cost for solid systems.
Another favorable strategy is to use the embedding scheme, which involves treating the strongly-correlated parts, such as the defects, with neural network QMC, while employing more cost-effective methods such as HF or DFT for other parts.

The flexibility of neural networks also enables modeling wavefunctions of multiple atomic configurations, which has been achieved for molecular systems and the applicability to solid systems is yet to be explored.
Specifically, neural networks for different systems can be simultaneously trained with a large portion of parameters shared, drastically reducing the total computational time compared to separate optimizations.~\cite{scherbela_deeperwin_2022}
An even more promising approach involves using a meta neural network to extract the atomic information and reparametrize the neural network for electron wavefunction.~\cite{gao_pesnet_2022,gao_sampling-free_2022,gao_generalizing_2023,scherbela_variational_2023,scherbela_towards_2024}
This approach not only saves computational time but also enables the development of a transferable wavefunction model. After pretraining this model on smaller fragments, it can be effectively applied to more extensive systems.
Moreover, it is feasible to pretrain a foundational wavefunction model, which can yield highly accurate energy results with only minimal computational effort.
Given their ability to handle diverse atomic configurations and the transferability to larger systems, neural networks hold great promise for performing more efficient finite-size error corrections and ensuring a reasonable computational for tasks requiring a substantial amount of atomic configurations, such as phonon calculation and force field training.

\section*{Funding Information}
JC is supported by the National Key R\&D Program of China under Grant No. 2021YFA1400500, the Strategic Priority Research Program of the Chinese Academy of Sciences under Grant No. XDB33000000, the National Natural Science Foundation of China under Grant No. 12334003, and the Beijing Municipal Natural Science Foundation under Grant No. JQ22001.

\selectlanguage{english}
\FloatBarrier
\bibliographystyle{vancouver}
\bibliography{main.bib}

\end{document}